 \documentclass[preprint2,epsf]{aastex}

\newcommand{\text}{\relax\ifmmode\let\next=\text@\else\let\next=\text@@\fi\next}
\newcommand{\mj}{M$_{\rm{J}}$}
\newcommand{\mc}{M$_{\rm{c}}$}
\newcommand{\rhc}{$\rho^1$\,Cnc}
\newcommand{\gae}{\mathrel{>\kern-1.0em\lower0.9ex\hbox{$\sim$}}}

\shorttitle{Upper Limit to Mass of $\rho^1$\,Cnc's RV Companion}
\shortauthors{McGrath et al.}

\begin{document}


\title{An Upper Limit to the Mass of $\rho^1$\,Cnc's Radial Velocity Companion}


\author{Melissa A. McGrath\altaffilmark{1}, Edmund Nelan\altaffilmark{1}, David. C. Black\altaffilmark{2}, George Gatewood\altaffilmark{3}, Keith Noll\altaffilmark{1}, Al Schultz\altaffilmark{1,4}, Stephen Lubow\altaffilmark{1}, Inwoo Han\altaffilmark{5}, Tomasz F. Stepinski\altaffilmark{2}, and Thomas Targett\altaffilmark{6}}

\altaffiltext{1}{Space Telescope Science Institute, 3700 San Martin Drive,
Baltimore, MD 21218; mcgrath@stsci.edu, nelan@stsci.edu, noll@stsci.edu, lubow@stsci.edu}
\altaffiltext{2}{Lunar and Planetary Institute, 3600 Bay Area Blvd,
Houston, TX 77058; black@lpi.usra.edu; tom@lpi.usra.edu}
\altaffiltext{3}{University of Pittsburgh, Allegheny Observatory, Observatory Station, Pittsburgh, PA 15214; gatewood@pitt.edu}
\altaffiltext{4}{Computer Sciences Corporation; schultz@stsci.edu}
\altaffiltext{5}{Korea Astronomy Observatory, Bohyunsan Optical Astronomy Observatory, Jacheon PO Box 1, Young-Chen, Kyung-Book, 771-820 Korea; iwhan@boao.re.kr}
\altaffiltext{6}{Cardiff University of Wales, Cardiff, CF24 3YB, Great Briton; TargettTA@cf.ac.uk}

\begin{abstract}
Doppler spectroscopy of \rhc\ has detected evidence of a companion with
an orbital period of 14.65 days and a minimum mass of 0.88 Jupiter
masses.  Astrometric observations performed with the Hubble Space
Telescope Fine Guidance Sensor 1r using a novel new observing technique
have placed an upper limit on the astrometric reflex motion of \rhc\ in
a time period of only one month. These observations detected no reflex
motion induced by the 14.65 day period radial velocity companion,
allowing us to place a $3\sigma$ upper limit of
$\sim$0.3\,milliarcseconds on the semi-major axis of this motion,
ruling out the preliminary Hipparcos value of 1.15\,milliarcseconds.
The corresponding upper limit on the true mass of the companion is
$\sim$30\,\mj, confirming that it is a sub-stellar object.
\end{abstract}



\section{Introduction}

The Doppler spectroscopy technique has been used in recent years to
detect low amplitude, periodic radial velocity variations in $\sim$60
nearby stars, which have been interpreted as due to planetary mass
companions.  This interpretation requires several assumptions, namely
that the root cause of the variation is Keplerian in nature; that the
companion mass (\mc) is substantially less than the mass of the
primary; and that one is seeing light from a single star as opposed to
an unresolved, comparable-mass binary system \citep{impr98}. The quantity determined by the radial velocity
observations is then \mc\,sin($i$) where $i$ is the unknown inclination of
the orbital plane to our line of sight to the star. The argument
usually advanced to suggest that the masses of the companions must be
small assumes that the distribution of orbital inclinations is uniform
so that, on average,
$\langle$\mc$\rangle\,=\,4\langle$\mc\,sin($i)\rangle/\pi$
\citep{chandra50}, which implies that the true companion mass cannot be
much greater than the minimum mass, \mc\,sin($i$).  Note that this
statement is only true for a large sample, and it does not preclude
individual companions from having true masses that are significantly 
larger than their minimum mass.

Several pieces of work support the planetary mass interpretation. For
example, \citet{itomay01} have determined that the dynamical stability
of the Ups And system constrains the mass of the outer planet to be
less than 1.43 times its minimum \mc\,sin($i$) value.  There is also
one case, HD\,209458, for which the companion has been observed to
transit the disk of the central star \citep{charb00}, hence giving a
measure of both the companion radius and the inclination of its orbit.
The companion mass of $\sim$0.6 Jupiter masses (\mj) is consistent with
that of a giant planet.  However, despite the statistical and dynamical
arguments, and the special case of HD\,209458, doubts have persisted
that \mc\,sin(i) is nearly equal to \mc\ for several reasons.

Using a combination of Hipparcos and ground-based MAP (Multichannel
Astrometric Photometer) astrometry in conjunction with the orbital
parameters derived from the radial velocity (RV) data, \citet{gwetal00}
have found that the \mc\,sin($i$) = 1\,\mj\ companion to $\rho$ CrB is
in a nearly face-on orbit with $i$ = 0.5\degr, and has a true mass of
115\,\mj, making it an M dwarf rather than a planetary-mass object.
\citet{zm00} have used Hipparcos astrometry to show that the
\mc\,sin($i$) = 6.4\,\mj\ companion to HD\,10697 is actually a
38\,\mj\ brown dwarf in an orbit with an inclination of just 5\degr.  A
similar analysis combining the Hipparcos and radial velocity data
extended to the 30 systems with orbital periods in excess of ten days
\citep{hanetal01} suggests that at least four of the 30 stars they
analyzed have stellar mass companions, that is, \mc\,$>\,80$\,\mj,
which in turn would require $i<1$\degr. If the distribution of
inclinations in the sample is uniform, the probability of a system
having an inclination $i\,<\,i_o$ is given by $1\,-\,$cos($i_o$).  For
$i_o$\,=\,1\degr, the probability would be 1.5$\times\,10^{-4}$, making
it unlikely that even one such system would be observed in a sample of
2000 stars. This led Han et al. to suggest that there might be a bias
toward small inclination angles in the radial velocity studies.
\citet{pour01} also finds statistically significant astrometric orbits
with low inclinations for three out of four of the Han et al.~potential
stellar mass companions. However, \citet{pourar01} argue that the trend
to low inclinations is an artifact of the adopted reduction procedure,
and that the astrometric data are not precise enough to allow the
conclusion that a significant fraction of the RV companions have
stellar masses.

An independent line of evidence also raises the possibility of stellar
mass companions for some of the 60 systems. Among a subset of 9 of the
stars of spectral type F with candidate planetary companions, Suchkov
\& Schultz (2001) have identified three potential binaries, HD\,19994,
HD\,89744, and HD\,169830, along with HD\,114762 (private
communication), based on the fact that they are overly bright for their spectral
type and Hipparcos distance. Gonzalez et al. (2001) found
that two stars with RV companions, HD\,37124 and HD\,46375, are similarly
too luminous for their spectral type and distance; they might each
be unresolved binary systems. However, they also note that a wide, long-period
binary cannot be ruled out, so the short period RV companion need not be the
source of the ``excess" luminosity.

Another line of reasoning suggesting that the RV companions may not be
planetary in nature comes from an analysis of the distribution of their
eccentricities and orbital periods, which are statistically
indistinguishable from those for single-line spectroscopic binaries
(SB1s) \citep{step01,heacox99}.  Moreover, the apparent correlation of
eccentricity with orbital period for the RV companions is similar to
that for SB1s \citep{black97,heacox99}, and their bivariate probability
distribution functions are again statistically indistinguishable
(Stepinski \& Black 2001). The observed orbital properties also differ
strongly from those of our own planetary system. The current data,
interpreted as planetary systems with random inclinations, also does not
lead to a simple theoretical picture (e.g., Marcy et al.~1999), and there
is currently no compelling dynamical argument to support the
interpretation that all of the RV companions are planets, rather than
brown dwarfs or stars in low inclination orbits. A determination of
which companions (if any) are not planetary would facilitate the
development of a dynamical model.

Astrometric observations of the stellar reflex motion induced
by a companion can potentially remove the uncertainty in sin($i$),
and thus determine the companion mass. The Hubble Space Telescope (HST) Fine Guidance Sensor 1r
(FGS1r) can measure relative stellar positions
to an accuracy $\sim$0.3\,milliarcsecond (hereafter mas), a
factor of 3-5 improvement over that of Hipparcos and MAP data. In this paper we present the results of a pilot
program utilizing an observing technique designed to quickly search for
perturbations larger than about 0.3\,mas to the position of
$\rho^1$\,Cnc (HR\,3522, HD\,75732, 55\,Cnc), a G8V star with an
\mc\,sin($i$) = 0.88\,\mj\ companion in an orbit with a period of 14.65
days \citep{betal97}. Han et al. report a preliminary reflex motion
with a semi-major axis of 1.15\,mas for this star. We discuss our
rationale for selecting this target and our observing strategy in
section 2, and our results in section 3.

\section{Observations}
Detection of companion-induced reflex motion is complicated by the need
to determine the typically much larger proper motion
($\mu_\alpha,\mu_\delta$) and parallax ($\pi$) of the star, which
normally requires observations over a baseline of at least one year.
However, a carefully designed observing strategy can shorten this time
dramatically provided an RV system with a relatively short period is
chosen, and provided the individual astrometric observations can be
made with sufficient precision. We therefore concentrated on stars
from Han et al.~with the following characteristics: relatively short
period; a large ($>1$\,mas) reflex motion inferred to be present with
at least moderate statistical significance; and ones most likely to
have companions that are M dwarfs or brown dwarfs (their Groups 2 and
3). An additional criterion was the suitability of the target for FGS1r
astrometry, namely, the availability and distribution of stars in the
FGS field of view, needed to define the inertial reference
frame. Using these criteria, $\rho^1$\,Cnc was determined to be the
most suitable target.  Its predicted reflex motion of 1.15\,mas radius
implies that the companion is an M-dwarf of 126\,\mj\ rather than the
\mc\,sin($i$) value of 0.88\,\mj. However, the Han et al.~result has
only borderline statistical significance for this target, and therefore
falls into their Group 2, stars for which they expect the majority of
the companions to be brown dwarfs.  

We realized that we would detect a reflex motion with FGS1r only if the
RV companion is more massive than about 40\,\mj\, which, to be
consistent with the radial velocity data, implies that the inclination
of its orbit would be nearly face on. (Inclinations larger than about
2\degr\ imply a companion mass that is too small to produce an
astrometric signature large enough to measure with HST.) Our observing
strategy was designed to optimize our ability to detect a perturbation
with the known 14.65\,day period and low eccentricity of the
companion's orbit, and the unknown value of $\Omega$ (the longitude of
the ascending node).

We performed a set of FGS1r observations in the one month period
centered around the time of maximum parallax factor in right ascension
(RA) (hereafter $\alpha_{max}$) on 1
May 2001.  We observed 
at pairs of epochs with times that were both phased with the
companion's 14.65 day period (P), and symmetric about the time of
$\alpha_{max}$. Explicitly, observations were executed in
pairs occurring at the same phase from April 17 through May 16. The
relative phases for each epoch are shown in Figure 1. With additional
observations on May 30 (epoch 10 at $+2$P), we monitored the star's
position over more than two full orbits of the companion.  The star's
proper motion $\mu_\alpha$ can then be determined from the epoch pairs
without a simultaneous determination of its parallax. Since the rate of
change of the parallax as projected along declination,
$d\pi_\delta/dt$, was constant to within 1\% at the time of
$\alpha_{max}$, it simply added to $\mu_\delta$ as a constant.
Measurable reflex motion would manifest itself as a larger {\it
dispersion} in the proper motions of the pairs. This strategy favors
shorter period systems since the reflex motion is then more apparent,
i.e., less diluted by the star's much larger proper motion.

Epochs 2, 4, 5, 6, and 8 each consisted of two HST visits, the other
epochs of 1 visit.  The data obtained in epoch 3 were severely degraded
due to a loss of lock on guide stars part way through the observing
sequence, so the proper motion measurement utilizing data from the
epochs 3 ($-{{1}\over{2}}$P) and 7 ($+{{1}\over{2}}$P) temporal pair
could not be made.  Each visit consisted of one HST orbit. All
observations were obtained at a fixed HST roll angle and pointing, and
the observing sequence for each visit consisted of 4--5 observations of
$\rho^1$\,Cnc interspersed with 1--4 observations of each of the 7
reference stars. One of the reference stars, $\rho$\,Cnc\,B, is the
proper motion companion to $\rho^1$\,Cnc.  Observations of this star
were unsuccessful in the first 6 epochs due to incorrect input
coordinates.

\section{Results}

In a given HST orbit the relative positions of the stars were
determined with a precision of about 1\,mas, as illustrated by the
residuals for \rhc\ shown in Figure 2.  Data from multiple epochs were
combined by a standard 6 parameter plate solution using the GAUSSFIT
program (Jeffereys, Fitzpatrick, \& McArthur 1987).  The solution
yields the proper motion of $\rho^1$\,Cnc and $\rho^1$\,Cnc\,B, and a
catalog of star positions in ($\xi$,$\eta$) space, along with residuals
to these quantities. When combining data from visits restricted to the
$\pm$ temporal pairs defined above, the parallactic displacement along
RA is not an issue since it is common to such data sets. However, when
other visits are included in the analysis, the parallax of
$\rho^1$\,Cnc must be accommodated in the solution. Since it was not
possible to solve for the parallax (because of the short interval over
which the observations were made), we adopted the Hipparcos value of
$\pi = 79.8$\,mas for $\rho^1$\,Cnc and $\rho$\,Cnc\,B, while
constraining the parallax and proper motion of the remaining reference
stars to zero. To determine our sensitivity to the choice of $\pi$,
solutions were done with $75 < \pi < 82$\,mas. The residuals in the
astrometric catalog produced for each assumed value of $\pi$ were
essentially constant, indicating that our particular choice of $\pi$
was not important provided it was correct to within a few percent.

The residuals of the star positions in the derived astrometric catalog
were small, 0.27--0.45\,mas for $\xi$, and 0.28--0.60\,mas for $\eta$,
(see Figure 2) with the exception of those for $\rho$\,Cnc\,B, which
were $\sim$1 mas, attributable to the fact that it was observed in only
5 of 15 HST orbits. The derived proper motions for $\rho^1$\,Cnc and
$\rho$\,Cnc\,B were ($\mu_{\alpha}$,$\mu_{\delta}$)\,=\,
($-515.2\pm6.6,-256.6\pm4.7)$ and ($-508.7\pm18,-221.7\pm12.6)$ mas/yr
respectively. The error bars include the cumulative 1-sigma errors
from all the fitted parameters. This compares with the Hipparcos values
of $-485.46\pm1.03,-234.40\pm0.72$ for $\rho^1$\,Cnc. The difference
between our proper motion and the Hipparcos value is most likely due to
unmodeled, but small, proper motion of one or more reference stars,
which can't be accurately measured over such a short time. We emphasize
that our technique and results do not depend on an accurate
determination of the absolute proper motion, but rather on the
differential accuracy with which we can determine it among epochs and
epoch pairs, so a systematic offset from the Hipparcos value is not relevant.

We used three different techniques to search for reflex
motion and to gauge the sensitivity of
our measurements.  First, we fixed the inclination at $i=0\degr$ and
the period of the companion to P=14.65\,days and solved for the phase and
radius of the reflex motion, along with residuals of these values,
using data from all 10 epochs. This solution yielded a reflex motion
with a radius of $0.08\pm0.2$\,mas, i.e., a non-detection.

Second, we again fixed the inclination and period of the companion,
then in simulations imposed a range of values for the phase angle and radius of the
reflex motion on the data, and solved for the phase and radius (again using
data from all 10 epochs) to assess how accurately we could recover them
if such perturbations were present.  The recovered values of
($r$,$\phi$) matched the imposed values to within the residuals
computed by the model for $r\gae0.3$\,mas. From this test we concluded
that our observations would have detected any 14.65\,day period
circular reflex motion with a radius in excess of 0.3\,mas at
1.5\,$\sigma$. No such motion was seen in the real data.
 
Our final and most sensitive test for detecting the reflex is to look
for dispersions in the proper motions computed from the $\pm$ temporal
pairs taken at epochs with identical parallax factor $\alpha$.  The
proper motion computed from epochs (1,9) and (3,7), each pair taken at
the same star/companion orbital phase, should agree and be the real
proper motion. Unfortunately, as mentioned earlier, the astrometry
obtained in epoch 3 were compromised, so only the data from epochs
(1,9) could be used to measure a ``reflex-free" proper motion. Table 1
(``Measured") shows the values of ($\mu_\alpha,\mu_\delta$) computed
from the three remaining $\pm$ temporal pairs. As expected, the average
value of ($\mu_\alpha,\mu_\delta$)=(516,251)\,mas/yr derived from the
pairs is very close to that derived using the data from all 10 epochs.
However, the dispersion of $\mu_\delta$ is nearly a factor of three
larger than that of $\mu_\alpha$, due primarily to the (1,9) pair,
which we attribute in part to the fact that the epoch 1 and 9
observations consisted of only a single HST orbit, while the epoch 2,
4, 6, and 8 observations consisted of two HST oribts each.

Note that the change of the star's position due to its orbit around the
star-companion barycenter has a position angle which is 180$\degr$
different in epochs (2,8) relative to that in epochs (4,6) (see Figure
1). Therefore any astrometric signature of the star's companion would
manifest itself most dramatically in the the proper motions computed
from these epoch pairs. The data from epochs (4,6) take on particular
significance because they span the shortest time interval among the
temporal pairs, hence any detectable reflex motion is least diluted by
the star's true proper motion computed from this pair.

To illustrate this, we conducted a series of simulations whereby we
again impressed upon the data a circular reflex motion with a 14.65 day
period, along with semi-major axes of 0.3, 0.5 and 1.0\,mas, and a
variety of phase angles. The results of these simulations are shown in
Table 1.  As expected, because it is the ``reflex free" pair, the
epochs (1,9) proper motion shows no change due to the imposed
perturbation, while the (4,6) pair has changed significantly, even for
the smallest 0.3\,mas perturbation.  In these simulations, a reflex
motion is absorbed into the proper motion computed from each $\pm$
temporal pair.  Therefore, the greater the reflex, the larger the {\it
dispersion} in the values of ($\mu_{\alpha},\mu_{\delta}$), even though
the perturbation has little effect on the average value of the proper
motion computed from all the temporal pairs. For example, the measured
dispersion in ($\mu_{\alpha},\mu_{\delta}$) is (5,14), while that from
the simulation with ($r,\phi$)\,=\,(0.3\,mas,103\degr) is (18,15) and
that from ($r,\phi$)\,=\,(0.3\,mas,13\degr) is (5,25).  We take the
{\it ratio} between the standard deviation of the measured proper
motion and the perturbed proper motion as a rough (and very
conservative) estimate of our ability to detect a real perturbation to the
proper motion. The standard deviations in Table 1 are computed in the
conventional fashion, i.e., as the square root of  ${\sum^{n=3}\,(value_i\,-\,average)^2}$ divided by $(n-1)$ using the
three epoch pair values listed in the Table for the three cases presented. For the 0.3, 0.5 and 1.0\,mas perturbations these ratios
(average for the two phases presented in Table 1) are 3, 4, and 7.
Comparison of the measurements and simulations for only the (2,8) and
(4,6) pairs would in fact be a more valid (and less conservative)
measure of the sensitivity because the (1,9) pair is unaffected by the
perturbation, and would give an even larger ratio.  We therefore
conclude that a reflex motion of 0.3\,mas is conservatively ruled out
at about the 3$\sigma$ level, and any reflex motion with a semi-major
axis in excess of 0.5\,mas is firmly ruled out at greater than
4$\sigma$.

In summary, the strategy we chose for attempting to astrometrically
detect and measure a reflex motion of $\rho^1$\,Cnc with HST/FGS1r
yielded the anticipated sensitivity. With just 14 orbits of HST time
over a 30 day interval we were able to conclusively rule out the
preliminary 1.15\,mas perturbation proposed by Han et al. Furthermore,
we ruled out at the 3$\sigma$ level a perturbation with an amplitude
greater than about 0.3\,mas.  This places an upper limit of
$\sim$30\,\mj\ on the mass of $\rho^1$\,Cnc's companion, implying that
it is a sub-stellar object.  We emphasize that this observing technique
is fully capable of detecting the larger and more statistically
significant astrometric perturbations for, e.g., $\rho$\,Cr\,B
and HD\,195019 found by Han et al. (2001) and Pourbaix
(2001) if they are real.
  
\acknowledgments
We thank our Program Coordinator,
Denise Taylor, and gratefully acknowledge the award of HST Director's Discretionary time for this
project. Support for this work was provided by NASA through contract NASW-4574 with the Universities Space Research Association, and
grant HST-GO-09240 to STScI, LPI, and the Univ.~of Pittsburgh
from the Space Telescope Science Institute, which is operated by the
Association of Universities for Research in Astronomy, Incorporated,
under NASA contract NAS5-26555.


\begin{table}
\begin{center}
\leftline{Table 1. Proper Motion Measurements and Simulations}
\begin{tabular}{rccc}
\noalign{\vskip 5pt}
\tableline\tableline
\noalign{\vskip 5pt}
\noalign{\leftline{Measurements [$\mu_{\alpha},\mu_{\delta}$ in units of mas/yr]}}
\noalign{\vskip 5pt}
\tableline
Epochs (1,9)& -520,-274 \\
       (2,8)& -509,-241 \\
       (4,6)& -520,-239 \\ 
Standard Deviation& 5,14 \\
\noalign{\vskip 5pt}
\tableline\tableline

\noalign{\vskip 10pt}
\noalign{\leftline{Simulation 1. Phase=103\degr (RA)}} 
Radius & 0.3& 0.5& 1.0 \\
\tableline
Epochs (1,9)& -520,-274& -520,-274& -519,-274 \\
       (2,8)& -501,-241& -497,-241& -485,-241 \\
       (4,6)& -551,-233& -571,-232& -621,-231  \\
Standard Deviation& 18,15& 27,15& 50,16 \\
\tableline\tableline

\noalign{\vskip 10pt}
\noalign{\leftline{Simulation 2. Phase=13\degr (Dec)}}
Radius & 0.3& 0.5& 1.0 \\
\tableline
Epochs (1,9)& -520,-274& -520,-274& -520,-274 \\
       (2,8)& -508,-248& -508,-253& -508,-265 \\
       (4,6)& -520,-203& -519,-183& -518,-132  \\
Standard Deviation& 5,25& 5,34& 5,56 \\
\tableline\tableline
\end{tabular}
\end{center}
\end{table}

\begin{figure}
\plotone{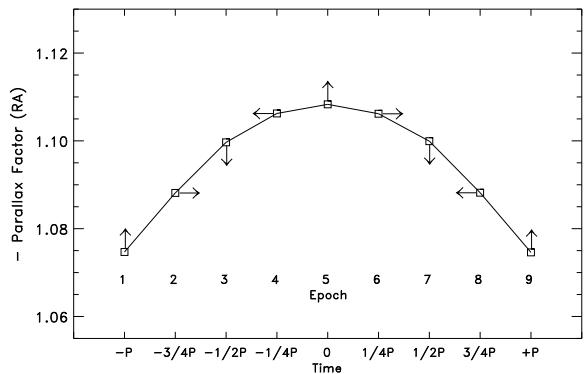}
\caption{The {\it relative} phase of observations at each epoch is illustrated schematically on this plot of parallax factor in RA vs time (in units of the RV companion period, P). Because we are searching specifically for a companion with a circular, face-on orbit with a period of 14.65 days, the absolute phase of the observations is not important. \label{fig1}}
\end{figure}

\begin{figure}
\plotone{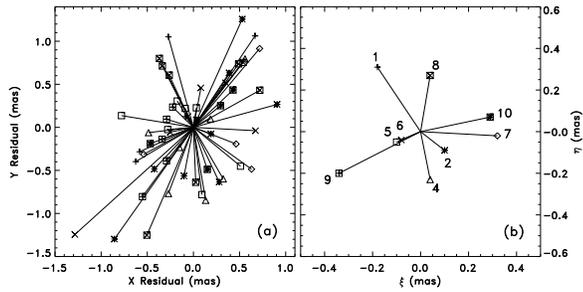}
\caption{(a) The X and Y residuals for all observations of \rhc\ relative to the best-fit 6 parameter plate solution model with proper motion and parallax
removed.  Data from the 10 epochs, excluding epoch 3, are plotted using symbols as in (b). (b) The $\xi,\,\eta$ position of \rhc\ at each epoch from the best-fit model, illustrating the accuracy of $\sim$0.3\,mas achieved by the observations.\label{fig2}}
\end{figure}


\begin{thebibliography}{}

\bibitem[Black(1997)]{black97} Black, D. C. 1997, \apj, 490, L171

\bibitem[Butler et al.(1997)]{betal97} Butler, R. P., Marcy, G. W., Williams, E., Hauser, H., and Shirts, P. 1997, \apj, 474, L115--118

\bibitem[Chandrasekhar \& M\"{u}nch(1950)]{chandra50} Chandrasaekhar, S., and
     M\"{u}nch, G. 1950, \apj, 111, 142--156

\bibitem[Charbonneau et al.(2000)]{charb00} Charbonneau, D., Brown, T. M.,
     Latham, D. W., and Mayor, M. 2000, \apj, 529, L45--L48

\bibitem[Gatewood, Han \& Black(2001)]{gwetal00} Gatewood, G., Han, I., and 
     Black, D. C. 2001, \apj, 548, L61-L63

\bibitem[Gonzales et al.(2001)]{gonz01} Gonzalez, G., Laws, C., Tyagi, S., and Reddy, B. E. 2001, \aj, 121, 432-452

\bibitem[Han, Black \& Gatewood(2001)]{hanetal01} Han, I., Black, D. C.,
     and Gatewood, G. 2001, ApJ, 548, L57-L60 

\bibitem[Heacox(1999)]{heacox99} Heacox, W. D. 1999, \apj, 526, 928

\bibitem[Imbert \& Pr\'{e}vot(1998)]{impr98} Imbert, M., and Pr\'{e}vot, L. 1998,
     \aap, 334, L37-L40

\bibitem[Ito \& Miyana(2001)]{itomay01} Ito, M. and Miyana, S. 2001, \apj, 552, 372--379

\bibitem[Jefferys et al.(1987)]{jef87} Jefferys, W., Fitzpatrick, M. J., and
McArthur, B. 1987, Celest Mech, 41, 39

\bibitem[Marcy et al.(1999)]{marcyetal99} Marcy, G. W., Butler, R. P., Vogt, S. S., Fischer, D., and Liu, M. C. 1999, ApJ, 520, 239--247

\bibitem[Pourbaix(2001)]{pour01} Pourbaix, D. 2001, \aap, 369, L22-L25

\bibitem[Pourbaix \& Arenou(2001)]{pourar01} Pourbaix, D., and Arenou, 
     F. 2001, \aap, in press

\bibitem[Stepinski \& Black(2001)]{step01} Stepinski, T. F., and Black, D.
     C. 2001, \aap, 371, 250--259

\bibitem[Suchkov \& Schultz(2001)]{such01} Suchkov, A. A., and Schultz, A.
     B. 2001, \apj, 549, L237-L240

\bibitem[Zucker \& Mazeh(2000)]{zm00} Zucker, S. and Mazeh, T. 2000, \apj, 
     531, L67--L69

\end{thebibliography}
\end{document}